\documentclass[preprint2]{aastex}
\usepackage{graphicx}
\usepackage{color}

\begin{document}

\title{Can China's FAST telescope detect extraterrestrial von-Neumann probes?}


\author{Osmanov Z.N.}
\affil{School of Physics, Free University of Tbilisi, 0183, Tbilisi,
Georgia}
\affil{E. Kharadze Georgian National Astrophysical Observatory, Abastumani, 0301, Georgia}
\email{z.osmanov@freeuni.edu.ge}

\begin{abstract}
In the present paper we consider the Type-2.x and Type-3.x extraterrestrial von-Neumann probes and study the problem of their detectability by the world's largest radio telescope: the Five-hundred-meter Aperture Spherical Radio Telescope (FAST). For this purpose we estimate the radio spectral parameters and analyse the obtained results in the context of technical characteristics of FAST. As a result, it is shown that FAST can detect as galactic as well as extragalactic self-replicating probes with high precision.
\end{abstract}


\section{Introduction}

Recently we have examined extraterrestrial von-Neumann probes and in the context of space colonisation efficiency we studied self-replicators' optimal sizes, reproduction time-scales, variability of brightness and spectral features \citep{probe2,probe1}. It has been shown that the replication time-scale, $\tau=\alpha r$, is proportional to the size of the probe, $r$, but by means of the interaction of the robots with the interstellar protons they accelerate and hence emit electromagnetic radiation. This process leads to the exponentially increasing luminosity of the swarm of probes behaving with $r$ as $\sim re^{t/(\alpha r)}$. On the other hand, it has been assumed that to colonize a certain area of space the luminosity should be limited either by the solar luminosity or by the galactic total luminosity. Therefore, imposing the aforementioned condition one can derive the optimal size of robots, which as it has been shown must be much less than millimetre. Then, characteristic timescales become several years, leading to potentially detectable flare type phenomena in a broad spectral band. Originally the idea of machines with artificial intelligence capable of self-replication has been proposed and studied by \cite{neum} where the author studied the problem in the context of information theory and thermodynamics. 

In \citep{probe2,probe1} we have considered Kardashev's Type-II and Type-III advanced civilizations. In the   framework of the mentioned classification Type-I is a technological civilization capable of harnessing the whole energy coming from the host star to a planet. Type-II and Type-III are advanced hypothetical societies using respectively the total energy of their host star and a host galaxy \citep{kardashev}. On the other hand, since an amount of energy harnessed by a civilisation in principle might be in between of amounts corresponding to Kardashev's Type-I and Type-II, or Type-II and Type-III, more appropriate designation is fractional \citep{Cirkovic}. Search for extraterrestrial intelligence (SETI) started from the last century, but after a significant contribution of \cite{dyson} the search for alien techno-signatures became one of the priorities of SETI and only last several years a series of works has been dedicated to this field \citep{paper1,paper2,paper3,Caplan,Haliki,Lacki,paper5}.

By \cite{probe2} it has been shown that Type-2.x,3.x von-Neumann probes might be visible also in the radio spectral band. In this context it is worth noting that the world's
biggest single-dish radio telescope, i.e., the China's Five-hundred-meter Aperture Spherical radio Telescope
(FAST), is dedicated to observe the sky in a frequency range $70$ MHz - $3$ GHz. Among others, one of the important missions of FAST is SETI project \citep{FAST} and recently the collaboration has announced the first SETI observations \citep{fseti,fseti1}. 
\begin{table}[ht]
\centering
\begin{center}
\begin{tabular}{ |c|c|c| } 
 \hline
 Illuminated aperture & 300 m\\ 
 Frequency range & 70 MHz-3 GHz\\ 
 System temperature & 30 K\\ 
 \hline
\end{tabular}
\end{center}

\begin{tabular}[t]{cc}

\end{tabular}
\caption{Technical characteristics of FAST}
\label{tab:caption}
\end{table}%

The aim of the present paper is to reexamine the results obtained by \cite{probe2,probe1} to consider the radio characteristics of extraterrestrial von-Neumann self-replicators and discuss the problem in the context of observations performed by FAST.

The paper is organized in the following way: in Sec. 2, we consider the spectral characteristics of Type-II and Type-III von-Neumann extraterrestrial self-reproducing probes and obtain corresponding signatures in the radio spectral band and analyse them in the context of technical characteristics of FAST and in Sec. 3 we outline the summary of the obtained results.

\section[]{Main consideration}

In this section we outline the results obtained in \citep{probe1}, estimate the emission characteristics of Type-2.x and Type-3.x von-Neumann extraterrestrial probes in the radio spectral band and study the possibility of their detection by the instruments of FAST.

During the motion of probes they encounter molecules 
and by collecting them - regularly replicate. Following the approach from \citep{probe1}, for spherical robots moving in a molecular cloud the replication time-scale is given by \footnote{In Eq. 1 by \cite{probe2} there is a typo: one should put a correct expression $\tau = \frac{4\xi}{3\beta}\times\frac{\rho}{m_0n}\times\frac{r}{c}$.}
$$\tau = \frac{4\xi}{3\beta}\times\frac{\rho}{m_0n}\times\frac{r}{c}\simeq 3.37\times \frac{\xi}{0.1}\times\frac{0.01}{\beta}\times$$
\begin{equation}
\label{tau} 
\;\;\;\;\times\frac{m_p}{m_0}\times \frac{\rho}{0.4\;g\;cm^{-3}}\times\frac{10^4cm^{-3}}{n}\times\frac{r}{0.1\;mm} \;yrs,
\end{equation}
where $\xi < 1$ is the fraction of the total volume filled with the material the probe is made of and it has been assumed that the free space inside the automata is bigger than the space filled with the material, $\rho$ denotes its density normalized by the density of Graphene which we consider as an example of the strongest and relatively lightest material), $\beta = \upsilon/c$, $\upsilon$ is the probe's velocity and $c$ denotes the speed of light and we have supposed that velocity is non-relativistic (to demonstrate that even in this scenario the process of replication is very efficient), $m_0$ is the encountering molecule's mass and $n$ is the number density of molecules in a corresponding media (nebula or galaxy), normalised by the typical value of hydrogen nebulae. As we have already mentioned in the previous section, we consider the probes moving with constant speed in space, collecting the interstellar material, when the interacting protons inevitably radiate. But in order to maintain constant speed, the total energy input in the swarm of replicators should be of the order of the emitted energy. On the other hand, energy input is the energy the civilization is able to utilize which in turn, should be limited by the level of technology the society belongs to. In particular, by \cite{probe2} it has been shown that the total luminosity of ensemble of robots is given by
\begin{equation}
\label{Lt1} 
L_{\rm tot}\simeq\frac{\pi}{3\kappa} nrce^2\beta^4N_0\times 2^{t/\tau},
\end{equation}
where $\kappa\leq 1$ is a dimensionless factor describing the acceleration distance and $N_0$ is the initial number of probes. For Type-2.x civilization the luminosity should not exceed the solar luminosity, $L_{\odot}\simeq 3.8\times 10^{33}$ erg s$^{-1}$ and for Type-3.x the maximum luminosity is limited by the value $3.6\times 10^{10}L_{\odot}$. Then the radius of a probe, to visit a $D = 2$ pc diameter cloud composed of hydrogen atoms ($m_0 = m_p$, where $m_p$ is the proton's mass), should be of the order of $0.013$ cm. The same value for Type-3.x alien society (aiming to visit a galaxy - with Milky Way parameters: $m_0 = m_p$, $n_0 = 1$ cm$^{-3}$) equals $0.025$ cm. Correspondingly, one can straightforwardly show that travelling time-scales, $t\simeq D/v$ are of the order of $653.7$ yrs (for Type-2.x) and $1.6\times 10^7$ yrs (for Type-3.x, $D = 50000$ pc) respectively with multiplication factors, $2^{t/\tau}$ being of the order of $8\times 10^{44}$ and $2\times 10^{58}$ respectively. It is worth noting that the optimal size depends on a media the replicators are going to colonise. For example, if the probes move in $H_2$-type clouds, the corresponding values are $0.026$ cm (Type-2.x) and $0.049$cm (Type-3.x). 

Generally speaking, it is assumed that the total energy that is used to maintain the motion of probes with almost constant velocity is harnessed with the maximum efficiency and for this purpose the swarm of probes should use a Dyson sphere (DS) built around a star. If the efficiency of the DS is not $100\%$, then it will be characterized by the waste radiation \citep{Wright}, which in turn might be detectable either in the infrared or in the visible spectral band, that might be an additional fingerprint. But it is clear that one of the important issues a civilisation should address is to minimise the waste energy and thus making its detection almost impossible. Another thing that we would like to emphasize is the process of energy transfer from a DS to a swarm, which might be performed either by means of some unknown mechanism or by using a beamed radiation. In the latter case, it is clear that the major purpose of the probes is to absorb the incident radiation as efficiently as possible. Consequently a reflected fraction of energy should be very low and therefore, it is expected that there is no significant reflected emission.

As it has been shown in \citep{probe2}, the radiated energy (by means of bremsstrahlung) per unit of frequency for a single probe is given by
\begin{equation}
\label{W2} 
\frac{dW}{df} = \frac{4\pi e^2\beta^2}{3c}\times\left(\frac{f_0}{f}\right)^2\sin^2\left(\frac{f}{f_0}\right),
\end{equation}
where $f$ is the radiated frequency and $f_0\equiv\beta c/(2\pi \kappa r)$ and we have taken into account $\omega\equiv 2\pi f$. By combining Eq.(\ref{W2}) with the incident proton flux, $nc\beta$, for the total spectral power one obtains
\begin{equation}
\label{W3?} 
\frac{dW}{dtdf} = 2^{t/\tau}n N_0\frac{4\pi e^2r^2\beta^3}{3}\times\left(\frac{f_0}{f}\right)^2\sin^2\left(\frac{f}{f_0}\right).
\end{equation}
The normalization frequency, $f_0$, for the typical parameters writes as follows
\begin{equation}
\label{freq} 
f_0\simeq\frac{\beta c}{2\pi r\kappa}\simeq  4.8\times 10^{10}\times\frac{0.1}{\kappa}\times\frac{\beta}{0.01}\times\frac{0.1\;mm}{r}\; Hz,
\end{equation}
where for the same reason as for $\xi$ we assumed $\kappa = 0.1$. From this expression\footnote{Unlike Paper-I, here we consider $\kappa = 0.1$.} it is evident that the ensemble of robots is characterised by broad spectra. Since the optimal sizes of self-replicators for Type-2.x and Type-3.x civilizations are respectively $0.26$ mm and $0.25$ mm, it is clear that for the radio frequency range the FAST is operating (see Table 1), the condition $f<<f_0$ is satisfied, which significantly simplifies Eq. (\ref{W3?}) to the following form
\begin{equation}
\label{W3} 
\frac{dW}{dtdf} \simeq 2^{t/\tau}n N_0\frac{4\pi e^2r^2\beta^3}{3},
\end{equation}
By considering Type-2.x von-Neumann probes colonising a spherical atomic nebula ($H$) with diameter $D = 2 $ pc, after taking into account $t\simeq D/\upsilon$, the radiation spectral power, $E_f\equiv dW/(dtdf)$, becomes 
$$E_{_{2.x,f}}\simeq 5\times 10^{23} \;erg\; s^{-1}\;Hz^{-1}\times$$
$$\times\frac{n}{10^4\;cm^{-3}}\times\frac{N_0}{100}\times\left(\frac{\beta}{0.01}\right)^3\times\left(\frac{r}{0.26\;mm}\right)^2\times$$
\begin{equation}
\label{W4} 
\;\;\;\;\;\;\;\times exp\left(\frac{D}{2\;pc}\times\frac{m_0}{2m_p}\times\frac{n}{10^4\;cm^{-3}}\times\frac{0.26\;mm}{r}\right).
\end{equation}
As it is clear from Table 1, the total radio frequency interval for FAST is $70$ MHz - $3$ GHz, therefore, by taking into account the bandwidth $\Delta f = 2.93$ GHz, the bolometric radio luminosity of Type-2.x probes is of the order of
\begin{equation}
\label{L2} 
L_{_{2.x}}\simeq E_{_{2.x,f}}\times\Delta f\simeq 1.5\times 10^{33}\; erg\; s^{-1}.
\end{equation}
Similarly, for the Type-3.x von-Neumann robots, one obtains the spectral power of the order of
\begin{equation}
\label{E3} 
E_{_{3.x,f}}\simeq 1.3\times 10^{33}\; erg\; s^{-1}\; Hz^{-1},
\end{equation}
leading to the bolometric radio luminosity
\begin{equation}
\label{L3} 
L_{_{3.x}}\simeq E_{_{3.x,f}}\times\Delta f\simeq4.0\times 10^{42}\; erg\; s^{-1},
\end{equation}
where we have taken into account that the Type-3.x civilization is colonising a whole galaxy, implying that $D = 50000$ pc (diameter of a galaxy) and $m_0 = m_p$, $n = 1$ cm$^{-3}$ (parameters of interstellar medium).  In Eq. (\ref{E3}) we do not show exact dependence on physical parameters, since it is already given in Eq. (\ref{W4}) but for different normalisation. One should emphasise that the aforementioned expressions are rather estimates than exact formulas, because FAST does not cover the whole range of $70$ MHz-$3$
GHz but instead, its facilities are designed to cover only portions. On the other hand, the whole range might be achieved by means of the ultra-wide band receivers developed for PARKES telescope\footnote{https://ieeexplore.ieee.org/document/7300180}

The illuminated aperture of FAST is $d = 300$ m (see Table 1), which means that the angular resolution of the telescope is given by
\begin{equation}
\label{res} 
\theta_m = 1.22\;\frac{\lambda}{d}\simeq 4.1\times 10^{-4}\times\frac{3\;GHz}{f}.
\end{equation}
Therefore, for the maximum distance, $\Lambda_{_{II}}$, where a length-scale, $D$, might be spatially resolved one writes
\begin{equation}
\label{dist1} 
\Lambda_{_{2.x}}\simeq \frac{D}{\theta_m}\simeq 4.9\times\frac{D}{2\;pc}\times\frac{f}{3\;GHz}\; kpc,
\end{equation}
where $D$ is normalised by the diameter of the molecular hydrogen nebula. From this expression it is clear that the value of $\Lambda_{_{2.x}}$ is minimum for the lowest resolution of the telescope, corresponding to frequency, $70$ MHz, and approximately equals $115$ pc. By taking into account the fact that in the Solar neighbourhood the Milky Way's surface mass density is $\Sigma\simeq (38 \pm 4)\;M_{\odot}$ pc$^{-2}$ \citep{dens}, then, the total mass of stars and stellar remnants in the area where the telescope can resolve $2pc$ scales is of the order 
$$M_{tot}\simeq\pi\Sigma \Lambda_{_{2.x}}^2\simeq $$
\begin{equation}
\label{mass} 
\simeq(2.9\pm 0.3)\times 10^9M_{\odot}\times\left(\frac{D}{2\;pc}\times\frac{f}{3\;GHz}\right)^2.
\end{equation}
By considering Type-3.x probes and setting a galactic length-scale, for the maximum resolving distance, $\Lambda_{_{3.x}}$, one obtains
\begin{equation}
\label{dist2} 
\Lambda_{_{3.x}}\simeq \frac{D}{\theta_m}\simeq 1.2\times 10^2\times\frac{D}{50\;kpc}\times\frac{f}{3\;GHz}\; Mpc.
\end{equation}
Taking a spherical volume inside the radius of the order of $120$ Mpc, one can straightforwardly show that the number of galaxies which might be spatially monitored by the instruments of FAST is approximately $10^7$ \citep{karach}.

Another important issue one has to address is the sensitivity of the telescope. The minimum spectral flux density that can be distinguished from the noise equals $kT_{sys}/A$, where $T_{sys} = 30$ K (see Table 1) is the so-called system temperature and $A \simeq 7\times 10^8$ cm$^2$ is the illuminated aperture area of the telescope. On he other hand, assuming the  isotropic source, the spectral flux density of the swarm of probes equals $E_f/(4\pi \Delta^2)$, where $\Delta$ is the maximum distance. By equating the aforementioned spectral flux densities for the maximum distance one obtains
\begin{equation}
\label{dist3} 
\Delta\simeq\sqrt{\frac{AE_f}{4\pi kT_{sys}}},
\end{equation}
which leads to the following values for Type-2.x
$$\Delta_{_{2.x}}\simeq 26.5 \;kpc\; \times$$
$$\times\left(\frac{n}{10^4\;cm^{-3}}\times\frac{N_0}{100}\right)^{1/2}\times\left(\frac{\beta}{0.01}\right)^{3/2}\times\frac{r}{0.26\;mm}\times$$
\begin{equation}
\label{dist4} 
\;\;\;\;\;\;\;\times exp\left(\frac{1}{2}\times\frac{D}{2\;pc}\times\frac{m_0}{2m_p}\times\frac{n}{10^4\;cm^{-3}}\times\frac{0.26\;mm}{r}\right),
\end{equation}
as well as Type-3.x civilizations
\begin{equation}
\label{dist5} 
\;\;\;\;\;\;\; \Delta_{_{3.x}}\simeq 1.3 \;Gpc.
\end{equation}

One should emphasize that Eq. (\ref{dist4}) is quite sensitive with the number density of interstellar molecules. On the other hand, Milky Way is not uniform and the aforementioned number density varies by up to $4$ orders of magnitude, which might lead to even higher values of maximum distances.

By considering the Milky Way surface area and the circle with radius $\Delta_{_{2.x}}$, one can estimate the common intersection area inside the Milky Way $S\simeq 1670$ kpc$^2$, where the probes can be detected. Here we have taken into account the distance from the Solar system to the center of the Galaxy, $8$ kpc. The obtained value is approximately $85\%$ of the total area of the Milky Way galaxy, implying that the sensitivity of FAST enables to cover almost the entire Galaxy. In particular, by assuming that the total number of stars in the Galaxy is $400$ billion, the instruments of FAST can cover the region containing almost $340$ billion stars. Concerning the extragalactic self-replicators, by assuming that throughout the Universe (with diameter $28.5$ Gpc) there are $\sim200$ billion galaxies, from Eq. (\ref{dist5}) one can directly estimate the number of galaxies which potentially might be covered by the FAST instruments: $150$ million.

From Eqs. (\ref{dist1},\ref{dist2},\ref{dist4},\ref{dist5}) it is clear that $\Lambda_{_{2.x}}<\Delta_{_{2.x}}$ and $\Lambda_{_{3.x}}<\Delta_{_{3.x}}$. This indicates that although the instruments might detect potentially interesting objects from very distant regions, they cannot be spatially resolved. Despite this fact, it does not mean that detection of flux on larger distance is useless. In particular, from Eqs. (\ref{tau},\ref{Lt1}) it is clear that the spectral power is characterised by the exponential growth with the corresponding timescale 
\begin{equation}
\label{tau1} 
\tau_{_{2.x}}^{\star} = \frac{\tau_{_{2.x}}}{\ln2}\simeq 6.3 \; yrs,
\end{equation}
where we have taken into account $m_0 = 2m_p$ and $r = 0.26$ mm in Eq. (\ref{tau}). From the analysis it is evident that there are two important signatures characterising Type-2.x von-Neumann probe candidates: (I) almost flat spectral power, independent of frequencies (see Eq. \ref{W3}) and (II) exponential growth of spectral power potentially detectable during several years.

After substituting $m_0 = m_p$, $n = 1$ cm$^{-3}$ and $r = 0.25$ mm in Eq. (\ref{tau}) and therefore, considering Type-3.x self-replicators, one arrives at the corresponding time scale of the spectral power exponential growth rate
\begin{equation}
\label{tau2} 
\tau_{_{3.x}}^{\star} = \frac{\tau_{_{3.x}}}{\ln2}\simeq 1.2\times 10^5 \; yrs.
\end{equation}
Despite the large variability time-scale, yet, all is not lost for those hoping to observe and identify extragalactic civilizations, i.e. their von Neumann probes. In particular, the exponential growth might lead to the detectable spectral power change in a reasonable time of observation. Likewise the previous case (see Eq. \ref{dist3}) the minimum change of the spectral power that can be distinguished from the noise is of the order of $kT_{sys}$, leading to the following maximum distance of detection
\begin{equation}
\label{dist6} 
\Delta^{\star}_{_{_{3.x}}}\simeq\sqrt{\frac{At_{obs}E_{_{3.x,f}}}{4\pi\tau_{_{3.x}}^{\star}kT_{sys}}}\simeq 4.0\times \left(\frac{t_{obs}}{1yr}\right)^{1/2}\; Mpc,
\end{equation}
where $t_{obs}$ denotes the time between two observations. As it is clear from Eq. (\ref{dist6}), the major emission fingerprints of candidate objects: flatness and independence of the radio spectrum on frequency and change of the value of $E_{_{3.x,f}}$ might be detectable for a quite large distance. Of course, the change of the spectral power does not carry the whole information of exponential growth rate, but still is detectable as a corresponding linear term.

\section{Conclusion}

We have considered the von-Neuman Type-2.x and Type-3.x extraterrestrial probes and studied the possibility of their detection by the instruments of FAST.

By analysing the emission spectral power we concluded that FAST can spatially resolve self-replicators on maximum distances of the order of $4.9$ kpc (Type-2.x) and $120$ Mpc (Type-3.x) respectively.

As it has been shown, corresponding maximum distance the flux of Type-2.x probes might be detected is of the order of $26.5$ kpc, covering almost $85\%$ of the whole Milky Way galaxy when observed from the Solar galactocentric distance. Up to these distances the exponential growth of luminosity can be measured with high precision.

Due to the large time-scale of growth of luminosity of Type-3.x self-replicators ($10^5$ yrs), the exponential growth cannot be directly detected, but the change corresponding to the exponential growth might still be measured up to the distances of the order of $4$ Mpc by two measurements shifted by $\sim 1$ year.

Finally, one can conclude that detection of flatness and independence of the spectral power on radio frequencies and high angular resolution of the instruments of FAST in the range $70$MHz-$3$GHz seems to be very prospective.

\section*{Acknowledgments}
I would like to thank an anonymous referee for valuable comments.


\begin{thebibliography}{99}
\bibitem[Bovy and Rix(2013)]{dens} Bovy, J. and Rix, H. 2013, ApJ, 779, 115
\bibitem[Caplan(2019)]{Caplan} Caplan, M.E. 2019, Acta Astr., 165, 96
\bibitem[Cirkovic(2015)]{Cirkovic} Cirkovic, M.M. 2015, SerAJ, 191, 6
\bibitem[Dyson(1960)]{dyson} Dyson, F., 1960, Science, 131, 1667
\bibitem[Haliki(2019)]{Haliki} Haliki, E., 2019,  IJAsB, 18, 455
\bibitem[Karachentsev(2005)]{karach} Karachentsev, I. D., 2005, AJ 129, 178
\bibitem[Kardashev(1964)]{kardashev} Kardashev, N. S., 1964, Sov. Astr., 8, 217
\bibitem[Lacki(2019)]{Lacki} Lacki, B.C., 2019, 131, 024102
\bibitem[Li et al.(2020)]{fseti1} Li, D. et al., 2020, Res. Astron. Astrophys., 20, 078
\bibitem[Lu et al.(2019)]{FAST} Lu, J., Lee, K. and Xu, R., 2019, Science China Physics, Mechanics and Astronomy, 63, 229531
\bibitem[von Neumann(1966)]{neum} Von Neumann, J., 1966, Theory of Self-Reproducing Automata 
(University of Illinois Press, Urbana and London)
\bibitem[Osmanov(2020a)]{probe2} Osmanov, Z., 2020a, JBIS, 73, 254
\bibitem[Osmanov(2020b)]{probe1} Osmanov, Z., 2020b, IJAsB, 19, 220
\bibitem[Osmanov(2017)]{paper2} Osmanov, Z., 2018, IJAsB, 17, 112
\bibitem[Osmanov(2016)]{paper1} Osmanov, Z., 2016, IJAsB, 15, 127
\bibitem[Osmanov and Berezhiani(2019)]{paper5} Osmanov, Z. and Berezhiani, V.I., 2019, JBIS, 72, 254
\bibitem[Osmanov and Berezhiani(2018)]{paper3} Osmanov, Z. and Berezhiani, V.I., 2018,
IJAsB, 17, 356
7
\bibitem[Wright(2020)]{Wright} Wright, J., 2020, SerAJ, 891, 174
\bibitem[Zhang et al.(2020)]{fseti} Zhang, Z. et al., 2020, ApJ, 891, 174

\end{thebibliography}
\end{document}